\documentclass[useAMS,usenatbib]{mn2e}

\usepackage{graphicx}
\usepackage{amsmath}
\usepackage[dvips,usenames]{color}
\def\aj{AJ}%
\def\apj{ApJ}%
\def\apjl{ApJ}%
\def\apjs{ApJS}%
\def\aap{A\&A}%
\def\aaps{A\&AS}%
\def\mnras{MNRAS}%
\def\pasp{PASP}%
\newcommand{\dmo}{\mbox{$(m\!-\!M)_{0}$}}

\newcommand{\av}{\mbox{$A_V$}}

\newcommand{\feh}{\mbox{\rm [{\rm Fe}/{\rm H}]}}

\newcommand{\Msun}{\mbox{$M_{\odot}$}}
\newcommand{\Teff}{\mbox{$T_{\rm eff}$}}

\newcommand{\comment}[1]{}
\newcommand{\beq}{\begin{equation}}
\newcommand{\eeq}{\end{equation}}
\newcommand{\beqa}{\begin{eqnarray}}
\newcommand{\eeqa}{\end{eqnarray}}

\newcommand{\fuw}{\mbox{${\rm F336W}$}}
\newcommand{\fvw}{\mbox{${\rm F475W}$}}
\newcommand{\fiw}{\mbox{${\rm F814W}$}}

\newcommand{\fva}{\mbox{${\rm F555W}_{\rm A}$}}
\newcommand{\fia}{\mbox{${\rm F814W}_{\rm A}$}}
\title[An eMSTO in NGC~411] {An extended main sequence turn-off in the
  Small Magellanic Cloud star cluster NGC~411\thanks{Based on
    observations with the NASA/ESA {\it Hubble Space Telescope},
    obtained at the Space Telescope Science Institute, which is
    operated by the Association of Universities for Research in
    Astronomy, Inc., under NASA contract NAS5-26555} }

\author[Girardi et al.]{L\'eo Girardi$^{1}$, Paul Goudfrooij$^{2}$,
  Jason S. Kalirai$^{2,3}$, Leandro Kerber$^{4}$, \newauthor Vera
  Kozhurina-Platais$^{2}$, Stefano Rubele$^{1}$, Alessandro
  Bressan$^{5}$, Rupali Chandar$^{6}$, \newauthor Paola Marigo$^{7}$,
  Imants Platais$^{8}$, Thomas H. Puzia$^{9}$
  \\
  $^{1}$ Osservatorio Astronomico di Padova -- INAF, 
  Vicolo dell'Osservatorio 5, I-35122 Padova, Italy \\
  $^{2}$ Space Telescope Science Institute, San Martin Drive,
  Baltimore, MD 21218, USA \\
  $^{3}$ Center for Astrophysical Sciences, 
  Johns Hopkins University, Baltimore, MD 21218, USA \\
  $^{4}$ Universidade Estadual de Santa Cruz, Rodovia Ilh\'eus-Itabuna, km
  16 -- 45662-000 Ilh\'eus, Bahia, Brazil \\
  $^{5}$ SISSA, via Bonomea 365, I-34136 Trieste, Italy \\
  $^{6}$ Department of Physics and Astronomy, The University of
  Toledo, 2801 West Bancroft Street, Toledo, OH 43606, USA \\
  $^{7}$ Dipartimento di Fisica e Astronomia Galileo Galilei,
  Universit\`a di Padova, Vicolo dell'Osservatorio 3, I-35122 Padova, Italy \\
  $^{8}$ Department of Physics and Astronomy, Johns Hopkins
  University, 3400 North Charles Street, Baltimore, MD 21218, USA \\
  $^{9}$ Department of Astronomy and Astrophysics, Pontificia
  Universidad Cat\'{o}lica de Chile, Avenida Vicu\~{n}a Mackenna 4860,
  Macul,\\ Santiago, Chile \\
}

\begin{document}

\date{To appear in MNRAS 
}

\pagerange{\pageref{firstpage}--\pageref{lastpage}} \pubyear{2013}

\maketitle

\label{firstpage}

\begin{abstract}
  Based on new observations with the Wide Field Camera 3 onboard the
  Hubble Space Telescope, we report the discovery of an extended main
  sequence turn-off (eMSTO) in the intermediate-age star cluster
  NGC~411. This is the second case of an eMSTO being identified in a
  star cluster belonging to the Small Magellanic Cloud (SMC), after
  NGC~419.  Despite the present masses of these two SMC clusters
  differ by a factor of $\sim\!4$, the comparison between their
  colour--magnitude diagrams (CMD) shows striking similarities,
  especially regarding the shape of their eMSTOs. The loci of main CMD
  features are so similar that they can be well described, in a first
  approximation, by the same mean metallicity, distance and
  extinction. NGC 411, however, presents merely a trace of secondary
  red clump as opposed to its prominent manifestation in NGC~419. This
  could be due either to the small number statistics in NGC~411, or by
  the star formation in NGC~419 having continued for $\sim\!60$~Myr
  longer than in NGC~411. Under the assumption that the eMSTOs are
  caused by different generations of stars at increasing age, both
  clusters are nearly coeval in their first episodes of star
  formation. The initial period of star formation, however, is
  slightly more marked in NGC~419 than in NGC~411.  We discuss these
  findings in the context of possible scenarios for the origin of
  eMSTOs.
\end{abstract}

\begin{keywords}
Stars: evolution -- 
Hertzsprung-Russell (HR) and C-M diagrams 
\end{keywords}

\section{Introduction}
\label{intro}

\begin{figure}
  \resizebox{\hsize}{!}{\includegraphics{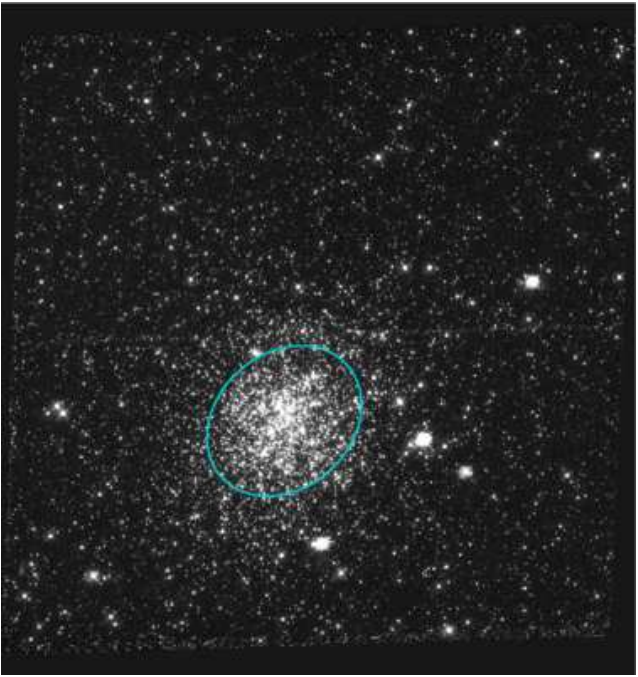}}
  \caption{The WFC3/UVIS image of NGC~411 in the filter F475W. The
    ellipse marks the cluster effective radius $R_{\rm eff}$. 
  }
\label{fig_image}
\end{figure}

The recent discovery of double, multiple, or simply extended main
sequence turn-offs (eMSTO) in intermediate-age Magellanic Clusters
\citep{Mackey_BrobyNielsen2007, Mackey_etal08} has generated much
interest in the literature, especially after its connection with the
phenomenon of multiple populations seen in old globular clusters
became clearer \citep{Goudfrooij_etal09, ConroySpergel11,
  Girardi_etal11, Keller_etal11, Mackey_etal13}. The present
understanding is that eMSTOs are simply reflecting multiple episodes
of star formation taking place inside the potential well of the
cluster, over a time span of a few 100~Myr.  Other effects such as a
spread in the rotational velocities \citep{BastiandeMink09} and the
presence of interacting binaries \citep{Yang_etal11} may contribute to
creating some moderate dispersion in the MSTO, but certainly cannot
cause the entire effect as observed: while the hypothesis of rotation
have been confuted on both theoretical and observational grounds
\citep[see][]{Girardi_etal00, Platais_etal12}, interacting binaries
are expected to make just a very minor fraction of the stars
populating such star clusters.

One of the most striking examples of eMSTOs is present in the massive
SMC cluster NGC~419 \citep{Glatt_etal09}, which also presents a dual
red clump (RC) of giants \citep{Girardi_etal09}. Detailed study of its
CMD by \citet{Rubele_etal11} has pointed to an internal age spread of
700~Myr. This is by far the longest detected duration of star
formation activity in a Magellanic Cloud star cluster. Massive LMC
clusters of the same age present star formation histories (SFH) which
last for at most 450~Myr \citep{Rubele_etal10, Rubele_etal13}.  Since
NGC~419 is the only known example of eMSTO in the SMC, it remains to
be explored whether this very long SFH is simply related with its
large mass, or if it has been favoured in some way by its host galaxy
-- either as a result of the different mean metallicity, or of the
different dynamical environment, of the SMC as compared to the LMC
disk.

In this paper, we present new HST observations that clearly show an
eMSTO in the relatively less massive SMC cluster NGC~411
(Sect~\ref{data}), which is nearly coeval to NGC~419
\citep{Rich_etal00}. Comparisons with theoretical isochrones and with
NGC~419 are presented in Sect.~\ref{sec_comp}.
Among other results, NGC~411 is found to have a more compact
RC than NGC~419, while sharing about the same spread of the turn-off.
Sect.~\ref{conclu} discusses these findings in the context of present
scenarios for the formation of eMSTOs.

\section{The NGC~411 data}
\label{data}

\subsection{Data and photometry}
\label{dataphot}

The NGC~411 data were obtained during HST cycle 18 program GO-12257
(PI: Girardi), and consist of total exposures of 2200, 1520, and 1980
seconds, respectively, in the \fuw, \fvw\ and \fiw\ filters of
WFC3/UVIS.  The cluster was centred slightly offset from the chip gap.
The latter was conveniently covered by means of two dithered
exposures. Fig.~\ref{fig_image} presents the WFC3 \fvw\ image
around the cluster centre.

A detailed description of the data reduction and analysis will be
provided in a forthcoming paper (P. Goudfrooij et al., in
prep.). Suffice it to mention that the images were processed
using standard techniques \citep[see][]{WFC3handbook} which include
bias and dark subtraction, cosmic-ray rejection, flat-fielding,
distortion correction and multi-drizzling. The photometry was
performed using two different methods: (1) the simultaneous ePSF
fitting technique developed by Anderson et al.~(2008), and later
adapted by him to be used in the WFC3/UVIS images, and (2) the
automated pipeline developed by J.\ Kalirai as described in
\citet{Kalirai_etal12}. Both reduction pipelines produced consistent
results, so that by default we will just present the data from method
(1). No correction for charge transfer inefficiency was performed.
The derived photometry was aperture-corrected and calibrated
into the Vegamag system as described in \citet{Goudfrooij_etal09}.

\subsection{Cluster structural parameters}
\label{parameters}

\begin{figure}
\resizebox{\hsize}{!}{\includegraphics{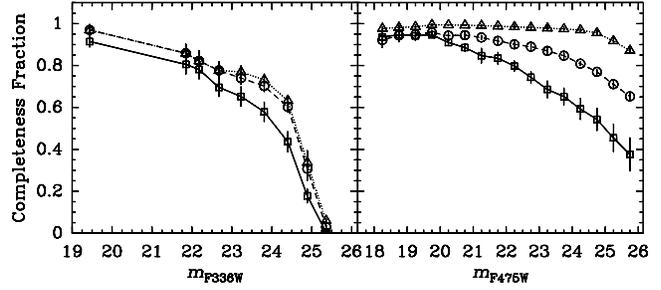}}
\caption{Completeness fraction as function of \fuw\ and \fvw\
  magnitudes and radius. The open squares and solid lines represent
  data within the core radius, the open circles and dashed lines
  represent data at a radius of 25\arcsec, and the open triangles and
  dotted lines represent data at a radius of 75\arcsec. Error bars
  depict the standard deviations among the independent runs of the
  artificial star tests.}
\label{fig_completeness}
\end{figure}

Extensive runs of artificial star tests were performed, giving origin
to the completeness functions illustrated in
Fig.~\ref{fig_completeness}, as a function of magnitude and of the
radius from the cluster centre.

\begin{figure}
\resizebox{\hsize}{!}{\includegraphics{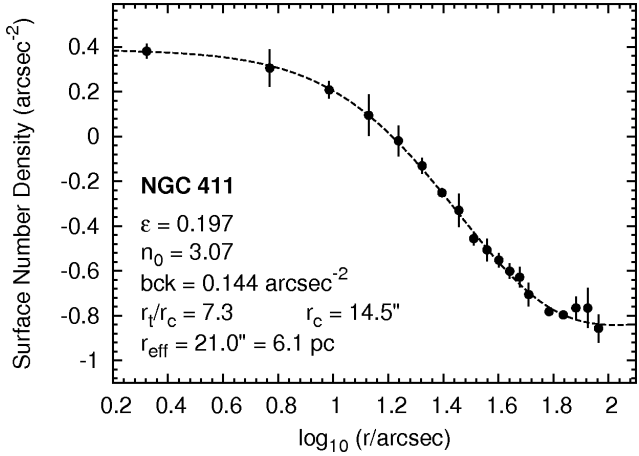}}
\caption{Radial surface density profile of NGC~411. The points and
  error bars represent observed values. The dashed line represents the
  best-fit \citet{King} profile whose parameters are shown in the legend.}
\label{fig_king}
\end{figure}

The best-fitting \citet{King} profile was derived using all stars with
$\fvw\!<\!23$~mag, and after applying completeness corrections, in the
same way as in \citet{Goudfrooij_etal11a}. The results are illustrated
in Fig.~\ref{fig_king}.
The cluster centre is located at coordinates $(x,y)=(2320, 2790)$ in
the drizzled image
(Fig.~\ref{fig_image}),
with a formal error of just 2 pixels in either direction.  Other
parameters for the fit are $R_{\rm c} = 14.5\pm0.9\arcsec$
($4.2\pm0.3$~pc), $R_{\rm eff}=21.0\pm1.6$~\arcsec ($6.1\pm0.5$~pc),
ellipticity $e=0.197$ and position angle ${\rm PA} = 48.4^\circ$.
These radii are geometric mean values, i.e.\ $R \equiv a\,\sqrt{1-e}$,
where $a$ is the semimajor axis of the ellipse.

For comparison, from the fits of King profiles
\citet{MackeyGilmore03smc} derive $R_{\rm c} =
9.72\pm0.36\arcsec=2.84\pm0.11$~pc, while \citet[][their tables 10 and
11]{McLaughlinvanderMarel05} derive $R_{\rm c}= 8.69\pm0.86\arcsec =
2.54\pm0.25$~pc and $R_{\rm h}= 21.9^{+1.2}_{-0.6}\arcsec =
6.40^{+0.35}_{-0.19}$~pc. In both cases $R_{\rm c}$ is much smaller
than our derived value.  Note that \citet{MackeyGilmore03smc} and
\citet{McLaughlinvanderMarel05} use the same set of surface brightness
data, originally obtained from \citet{MackeyGilmore03smc} from the
archival WFPC2 images. The smaller area and depth covered by WFPC2
observations, together with overlooking the apparent cluster's
ellipticity, might be at the origin of these different $R_{\rm c}$
values.

Throughout this paper, NGC~411 will be compared to NGC~419, which
apparently is a much bigger cluster in several aspects. According to
\citet{Glatt_etal09}, NGC~419 has $R_{\rm c}\! =\! 15.6$\arcsec, and
$R_{\rm eff}\!=\!35$\arcsec. Note that NGC~411 turns out to have about
the same $R_{\rm c}$ as NGC~419, but a much smaller $R_{\rm eff}$ (and
half-light radius).  NGC~411 and NGC~419 have integrated $V$-band
magnitudes, inside aperture radii of 100\arcsec, of 11.806 and
10.304~mag, respectively \citep{Goudfrooij_etal06}. Under the
assumption that both clusters have similar distances, extinction,
ages, metallicities (see below) and mass functions, this magnitude
difference translates into a ratio between the present total
masses\footnote{The integrated 2MASS $J$-band magnitude from
  \citet{Pessev_etal06} produces a luminosity ratio of 0.21. This
  estimate is however expected to be more affected by stochastic
  fluctuations due to the small numbers of bright RGB and AGB stars
  per cluster, so we opt for using the $V$-band one.} of $M_{\rm
  T}^{\rm NGC\,411} \simeq 0.25 \times M_{\rm T}^{\rm NGC\,419}$.

\subsection{Colour magnitude diagrams}
\label{cmds}

\begin{figure*}
\resizebox{0.83\hsize}{!}{\includegraphics{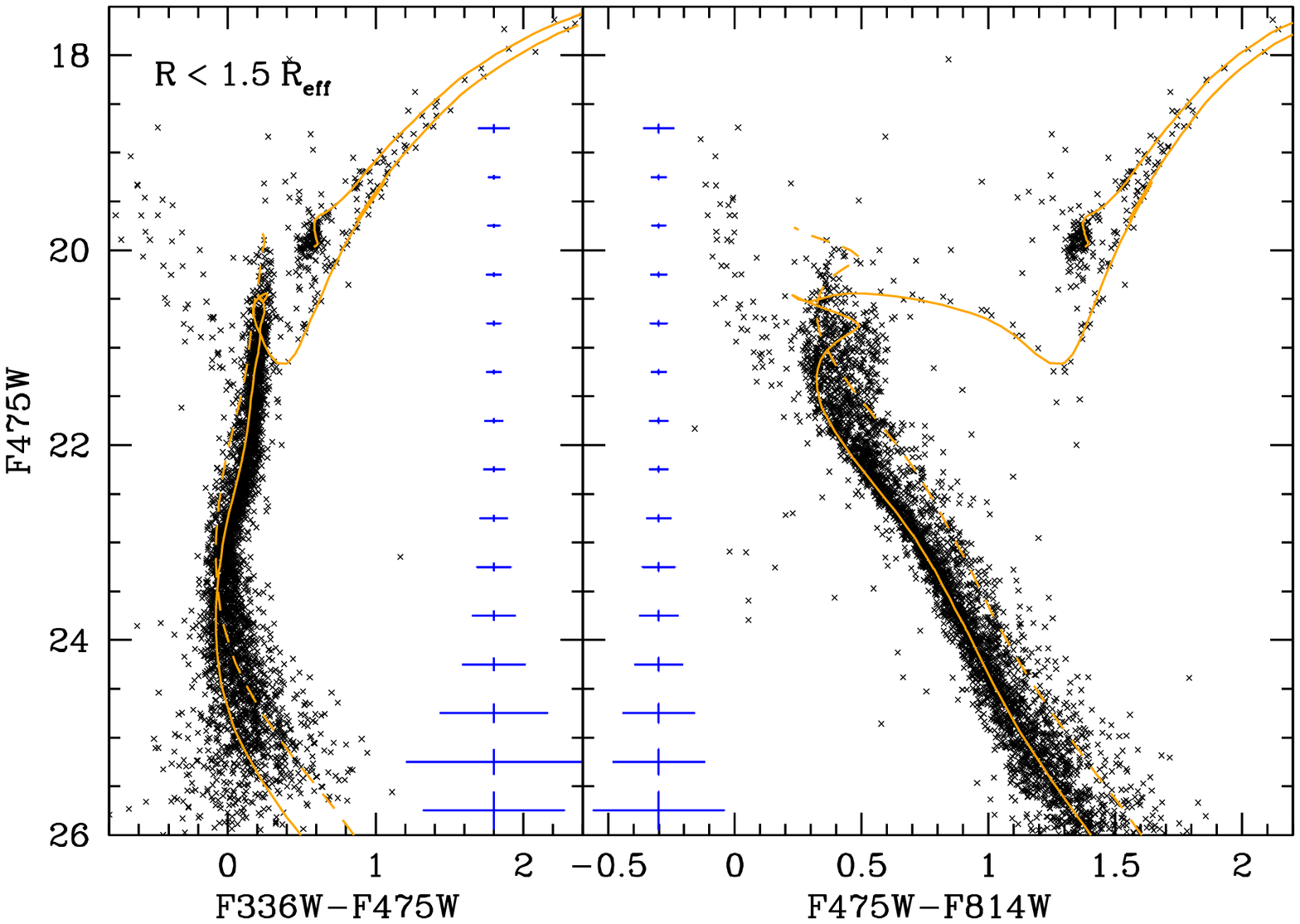}}
\resizebox{0.83\hsize}{!}{\includegraphics{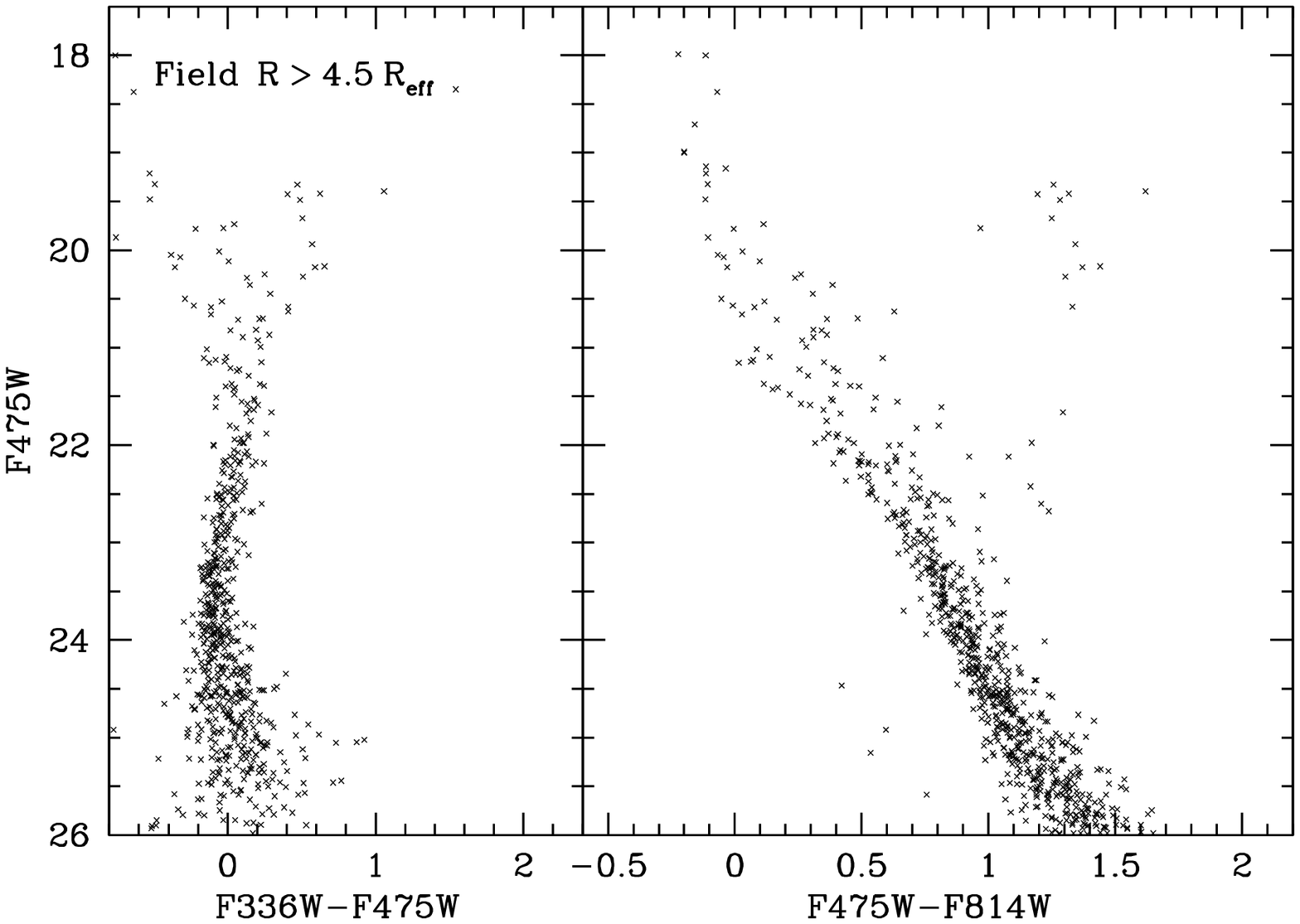}}
\caption{\textbf{Top panels:} \fvw~vs.~\fuw$\!-\!$\fvw\ and \fvw\ vs.\
  \fvw$\!-\!$\fiw\ diagrams for stars inside a radius of $1.5\, R_{\rm
    eff}$ of the NGC~411 centre (dots). The continuous orange line is a
  1.66-Gyr $\feh\!=\!-0.8$ isochrone shifted by $(\dmo,\av)=(18.9,
  0.25)$. The dashed line indicatively illustrates the expected
  sequence of equal-mass main sequence binaries for this same
  isochrone. Typical photometric errors are indicated by the blue
  error bars. \textbf{Bottom panels:} the same CMDs for a subsample of
  the field stars, randomly selected so as to represent the same area
  as in the top panels. }
\label{fig_cmds}
\end{figure*}

Figure~\ref{fig_cmds} shows CMDs derived from our data. The top panels
refer to the $0.87$~arcmin$^2$ region inside the limit of
$R<1.5\,R_{\rm eff}$, which is prevalently populated by cluster stars.
In order to illustrate the contamination by the field, the bottom
panels show a random subsample ($68$~\%) of the field stars at
$R>4.5\,R_{\rm eff}$, aimed to represent the same total area as the
top panels.  It is evident that the field contamination is close to
negligible in the cluster core.  This can be quantified by star counts
in the RC: the cluster $R\!<\!1.5\,R_{\rm eff}$ region contains 158 RC
stars (here broadly defined as stars with $19.5\!<\fvw\!<20.5$,
$1.2\!>\!\fvw\!-\!\fiw\!>\!1.5$), whereas the same area of the field
contains just $\sim\!5$ RC stars.

A feature evident in the new data is the broad turn-off of NGC~411,
especially noticeably in the \fvw\ vs.\ $\fvw\!-\!\fiw$ CMD of
Fig.~\ref{fig_cmds}.  The \fvw vs.\ $\fuw\!-\!\fvw$ CMD does not add
much to the resolution of this feature, due to the limited colour
range at which the turn-offs appear in this diagram.

The eMSTO of NGC~411 has not been noticed in the previous WFPC2 data
from \citet{Rich_etal00}, probably because of the much higher
photometric errors at that time. In our observations, the photometric
errors at the turn-off level (\fvw$\sim\!21$) are just 0.03~mag,
which is small enough to resolve these fine CMD structures.

In addition to the eMSTOs the cluster CMDs show other well-known
features, such as the sequence of binaries parallel to the main
sequence, the red giant branch (RGB), a well defined sequence of
subgiants, and the early-asymptotic giant branch bump. These sequences
will be discussed in more detail in a subsequent paper.

\section{Discussion}
\label{sec_comp}

\subsection{Comparison with isochrones}

We now give a closer look at the CMD features in NGC~411. Given the
insignificance of the field contamination, we will use just the stars
at $R\!<\!1.5\,R_{\rm eff}$. These CMDs will be compared to
theoretical isochrones, and to the NGC~419 data, for which the SMC
field contamination is also not a problem \citep{Girardi_etal09}.

\begin{figure}
\resizebox{\hsize}{!}{\includegraphics{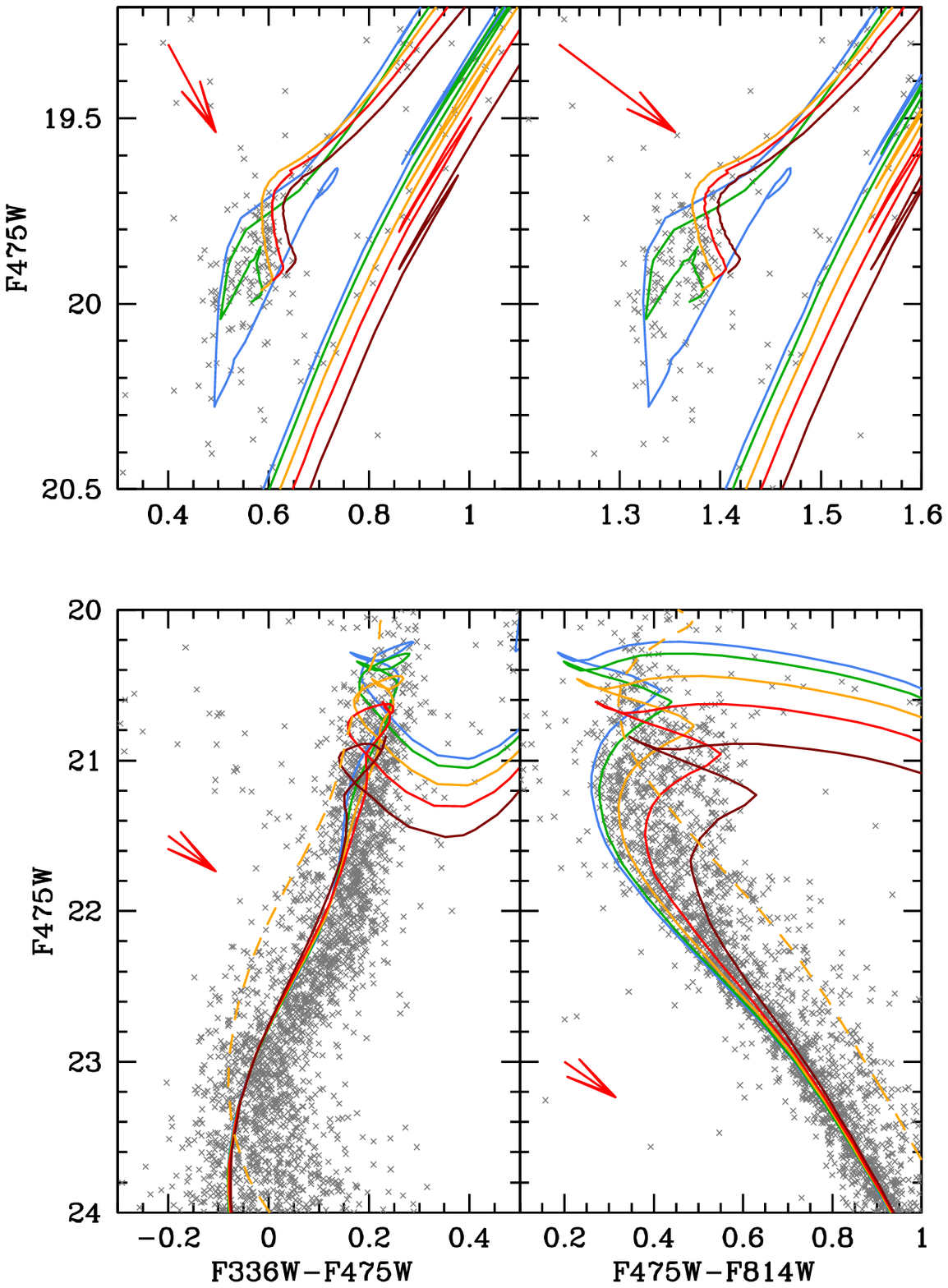}}
\caption{The same as in the top panels of Fig.~\ref{fig_cmds}, but
  zooming into the MSTO (bottom panels) and RC (top) regions of the
  CMDs. The extended structure of the MSTO and RC are readily seen.
  The red arrows indicate the reddening vector always for the same
  total extinction of $\av=0.2$~mag. Isochrones for ages 1.44, 1.51,
  1.66, 1.86 and 2.19 Gyr are overplotted (see text for more
  details).}
\label{fig_isoc}
\end{figure}

Figure~\ref{fig_isoc} zooms in the MSTO and RC regions in the \fvw\
vs.\ \fuw$\!-\!$\fvw\ and \fvw\ vs.\ \fvw$\!-\!$\fiw\ diagrams of
NGC~411, comparing the cluster data with theoretical isochrones
selected from the {\sc PARSEC} suite v1.1
\citep{parsec}\footnote{http://stev.oapd.inaf.it/cmd; {\sc PARSEC}
  isochrones are revised versions of previous Padova sets,
  converted to the HST WFC3 and ACS filters using the set of
  transformations described in \citep{Girardi_etal08}. }. The
isochrones illustrated are basically an eyeball fit to the \fvw\ vs.\
\fvw$\!-\!$\fiw\ diagrams on the right panel, with the \fvw\ vs.\
\fuw$\!-\!$\fvw\ on the left panel being used as a consistency check.
Five different ages are illustrated, as described below. One of these
isochrones is also overplotted in the previous Fig.~\ref{fig_cmds}.

We initially select isochrones with a metallicity of
$\feh\!=\!-0.8$~dex ($Z\! = \!0.00244,Y\! = \!0.25316$) and a
scaled-solar mixture of metals. With this choice, the position of the
main sequence (for $\fvw\!\ga\!22$~mag) is determined by two parameters
only: the absolute distance modulus \dmo, which just shifts the
isochrones vertically in the plot, and the total extinction \av, which
moves the isochrones along the extinction/reddening vectors
illustrated in the figure (for $\av\!=\!0.2$~mag). A good fit of the
observed MS in \fvw\ vs.\ \fvw$\!-\!$\fiw\ diagram -- and in
particular of its leftmost ridge line, which corresponds to sequence
of single hydrogen-burning stars -- is provided by the pair of values
$(\dmo,\av)=(18.9, 0.25)$. The rightmost sequence of stars along the
MS is easily explained as being a sequence of binaries (both apparent
and real). Indeed, the expected position of equal-mass detached
binaries is illustrated by the dashed line in Figs.~\ref{fig_cmds} and
\ref{fig_isoc}, which is obtained by simply shifting the MS section of
the isochrone upward by $-2.5\log(2)$~mag.

The same $(\dmo,\av)$ values provide also a good description of the MS
in the \fvw\ vs.\ \fuw$\!-\!$\fvw\ diagram, but offset in colour by
about $-0.05$~mag. We verified that it is not possible to get rid of
this offset by simply changing the isochrone metallicity by
$\pm0.2$~dex and hence adjusting the $(\dmo,\av)$ values.
Rather, we attribute the offset to a possible problem in the
colour-\Teff\ relations adopted in the isochrones. This question is
irrelevant for this work, since offsets of $\sim\!0.05$~mag in this
diagram would not change any of our conclusions.

After the metallicity and $(\dmo,\av)$ values are fixed, the age range
of the isochrones is defined by the location and width of the ``golf
club'' that characterises the eMSTOs in the \fvw\ vs.\
\fvw$\!-\!$\fiw\ diagram. We find that the age range from 1.51 to
2.19~Gyr ($\log(t/{\rm yr})$ from 9.18 to 9.34) comprises the extent
of the observed eMSTO in NGC~411. This $0.68$~Gyr interval should be
considered just as a rough estimate, since a more accurate evaluation
requires taking into account, at least, the photometric errors and the
spread in the CMD caused by binaries. Among the isochrones in this age
interval, the 1.66-Gyr one ($\log(t/{\rm yr})=9.22$, in orange)
reproduces particularly well the mean location of the subgiant branch,
and hence is taken as the representative ``mean age'', and is also
plotted in Fig.~\ref{fig_cmds}.

The selected age values are also supported by the comparison between
these theoretical isochrones and the location of the RC, shown in the
upper panels of Fig.~\ref{fig_isoc}. Most of the RC stars appear in a
compact and almost-vertical structure towards the red side of the RC.
This main concentration is expected for He-burning stars derived from
low-mass stars -- i.e.\ stars that developed an electron-degenerate
core after the MS and hence climb the RGB, igniting helium at a core
mass of $\ga0.46$~\Msun\ (see figure 10 in \citealt{parsec}).
Isochrones for all ages $\ga\!1.6$~Gyr are able to fit this rightmost
RC feature, which comprises about $2/3$ of the RC stars seen in
NGC~411.

The remaining RC stars are scattered in a lopsided feature towards the
blue and the faint part of the CMD, but still confined over a small
section of the \fvw\ vs.\ \fvw$\!-\!$\fiw\ diagram (spanning just
$\sim\!0.08$~mag in colour). We note that the isochrone of age
1.51~Gyr (in green) comprises the bulk of stars in this CMD feature.
In the isochrones, this feature clearly corresponds to the initial
section of the so-called ``secondary red clump'' \citep[SRC;
][]{Girardi99}, which is the youngest part of the RC, made of stars
just massive enough to have ignited helium in non-degenerate
conditions.

\subsection{Comparison with NGC~419}

We now compare the NGC~411 data and best-fit isochrones with the data
for the NGC~419 cluster which, as first noticed by
\citet{Rich_etal00}, has about the same age as NGC~411. NGC~419 is
also a cluster with a well-marked eMSTO \citep{Glatt_etal09} and a
dual red clump \citep{Girardi_etal09}. When interpreted in terms of
age spread only, these CMD features lead to an impressive
$\sim\!700$-Myr long period of star formation in NGC~419
\citep{Rubele_etal10}.

\begin{figure}
  \resizebox{\hsize}{!}{\includegraphics{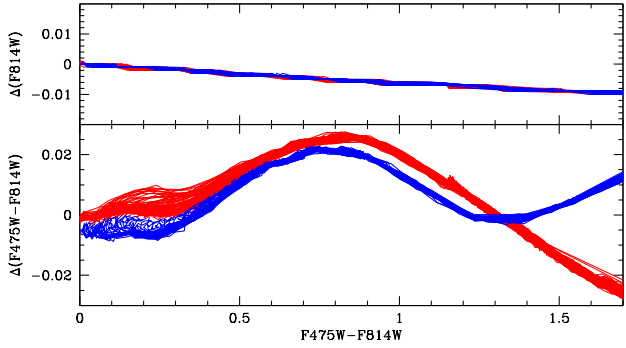}}
  \caption{The difference between colours and magnitudes in the WFC3
    filters system, as estimated from ACS colours and magnitudes via
    Eqs.~\ref{eq_trans}, and the values on the WFPC3 system, as a
    function of $\fvw\!-\!\fiw$ (see Eq.~\ref{eq_diff}).  These
    relations are derived from $\feh\!=\!-0.8$ model isochrones
    covering the age range from 1 to 4 Gyr, and separately for dwarfs
    ($\log g>4$; blue) and giants ($\log g<4$; red).}
\label{fig_trans}
\end{figure}

NGC~419 data comes partially from the ACS/WFC observations in GO-10396
(PI: J.S.\ Gallagher) in the filters \fva\ and \fia, re-reduced in a
similar way as the NGC~411 data\footnote{Throughout this paper, we
  denote ACS/WFC filters with the subscript A. When no subscript is
  provided, we are referring to WFC3/UVIS filters}. To these ACS
images, we have added \fuw\ data from WFC3/UVIS obtained in our
GO-12257 program, with the cluster centre positioned just slightly
offset from the chip gap as for NGC~411.  Since the filters are not
the same as for NGC~411, we have derived the equivalent of WFC3
filters using the following approximations derived from \citet{parsec}
isochrones\footnote{The isochrones take fully into account the
  different filter throughputs of ACS/WFC and WFC3/UVIS filters, and
  their impact on bolometric corrections and \Teff-colour
  transformations \citep[see][]{Girardi_etal08}.}:
\begin{eqnarray}
  \fiw&\simeq&\fia \nonumber \\
  \fvw\!-\!\fiw&\simeq&1.43\,(\fva\!-\!\fia) \,\,\,.
  \label{eq_trans}
\end{eqnarray}
These approximations give origin to the quantities
\begin{eqnarray}
  \Delta(\fiw)&=&\fia-\fiw \nonumber \\
  \Delta(\fvw\!-\!\fiw)&=&1.43\,(\fva\!-\!\fia)
  \nonumber\\ & & -(\fvw\!-\!\fiw)\,\,\,,
  \label{eq_diff}
\end{eqnarray}
which are plotted in Fig.~\ref{fig_trans} as a function of the colour
in the WFC3/UVIS system, i.e.\ $\fvw\!-\!\fiw$. As can be appreciated,
the relations from Eq.~\ref{eq_trans} turn out to be accurate to
within 0.01~mag in magnitude, and 0.02~mag in colour, over the entire
$0\!<\!(\fvw\!-\!\fiw)\!<\!1.5$ interval of interest in this paper,
and for both dwarfs and giants. We checked that the same numbers apply
over the entire metallicity range $-1.5\!<\!\feh\!<\!-0.5$, which
comprises most of the metallicity determinations of SMC field and
cluster stars \citep[e.g.][]{Carrera_etal08, Piatti12}. Such
$\la0.02$~mag errors, though systematic, are small enough to be
ignored in the context of this paper.

\begin{figure}
\resizebox{\hsize}{!}{\includegraphics{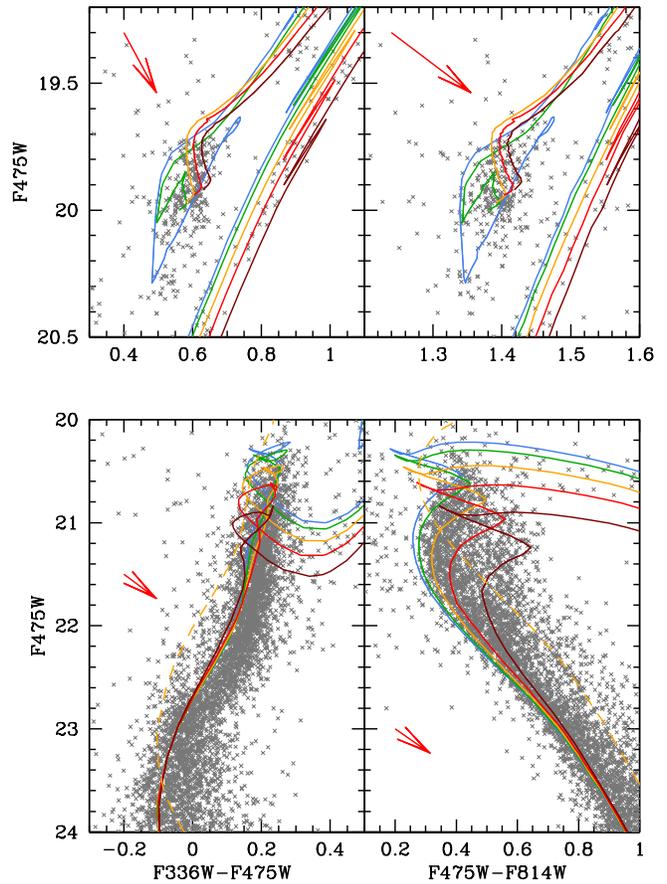}}
\caption{The same as in Fig.~\ref{fig_isoc}, but now for the cluster
  NGC~419. The data has been converted to the same scale and filters
  as NGC~411 by means of Eq.~\ref{eq_trans}. The isochrones are for
  exactly the same parameters as in Fig.~\ref{fig_isoc}.}
\label{fig_419}
\end{figure}

NGC~419 is located in an area towards the SMC Wing with about 2.3
times more SMC field population than that for NGC~411.  Moreover, the
NGC~419 centre presents a much larger density and is not as free from
crowding as NGC~411.  In order to circumvent these problems, we have
selected NGC~419 stars within an annulus with radius between 400 and
800 WFC3/UVIS pixels\footnote{We recall that 1 ACS/WFC pixel
  corresponds to 1.25 WFC3/UVIS pixel.} (16 to 32\arcsec, or
$\simeq4.7$ to 9.3 pc) from the cluster centre. This represents an
area comparable to the one selected for NGC~411, but with about twice
its numbers of cluster and SMC field stars.

Figure~\ref{fig_419} presents the comparison between NGC~419 data and
the same theoretical isochrones used in Fig.~\ref{fig_isoc}. Isochrone
parameters -- including distance, extinction, metallicity and ages --
are exactly the same as used for NGC~411.  A difference w.r.t.\
Fig.~\ref{fig_isoc}, is that the isochrone colours and magnitudes,
presented in the WFC3 filters, were also derived from the ACS ones via
the relations in Eq.~\ref{eq_trans}. In this way, we keep a high level
of consistency between data and models, while producing a plot that
has a very similar colour-magnitude scale as the one for NGC~411.

\begin{figure}
\resizebox{\hsize}{!}{\includegraphics{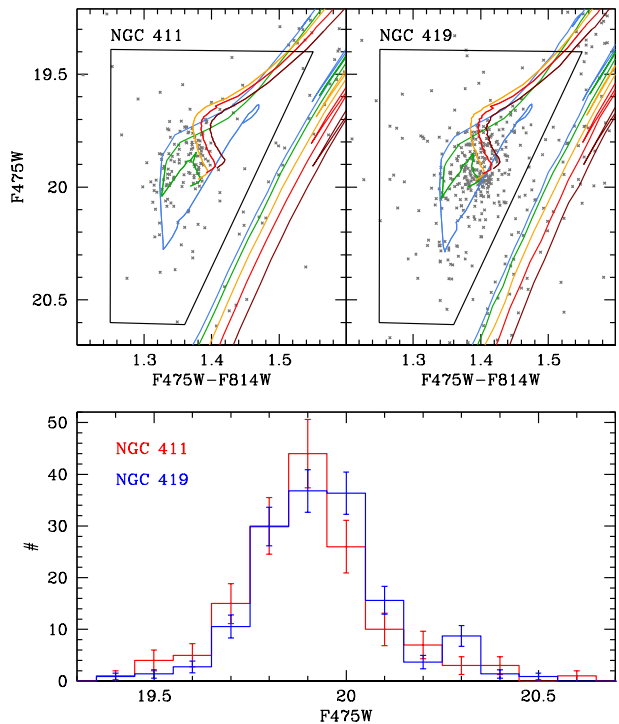}}
\caption{The top panels detail the selection of cluster stars in the
  RC region of both NGC~411 and NGC~419, inside the region delimited
  by black lines, which is designed to include as much RC as possible,
  but avoiding the sampling of first-ascent RGB stars. A total of 149
  RC stars are found inside the NGC~411 $1.5\,R_{\rm eff}$; they give
  origin to the red LF in the bottom panel. The same CMD region in
  NGC~419 contains 324 stars.  The NGC~419 LF has been scaled down so
  as to present the same total number of RC stars as NGC~411. The
  error bars are simple 1$\sigma$ error estimates, based on the
  square-root of the observed numbers.}
\label{fig_lfrc}
\end{figure}

\begin{figure}
\resizebox{\hsize}{!}{\includegraphics{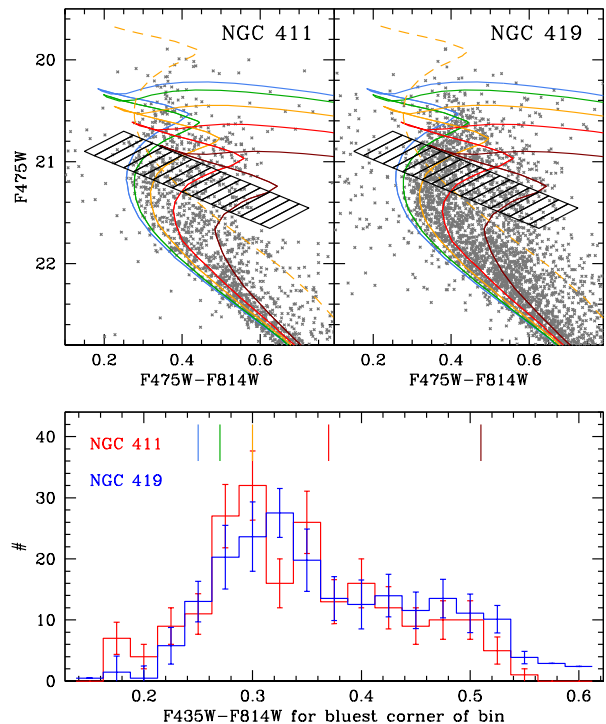}}
\caption{The top panels detail the selection of cluster stars across
  the eMSTOs of both NGC~411 and NGC~419, along a sequence that goes
  from left to right, inside the inclined parallelogram. Stars are
  counted inside the small boxes, which run almost parallel to
  the isochrones drawn for these clusters (the same as in
  Fig.~\ref{fig_isoc}). A total of 208 stars are found along this
  sequence in NGC~411; they give origin to the red histogram in the
  bottom panel. The abscissa in this case is simply the colour at the
  bottom-left extremity of the small boxes. For NGC~419, 425 stars are
  found inside the same CMD region; their histogram is plotted in the
  bottom panel (blue histogram) after being rescaled using a factor of
  0.49 to equalize the total number of stars in both clusters.  The
  error bars are simple 1$\sigma$ estimates, based on the square-root
  of the observed numbers.}
\label{fig_lfms}
\end{figure}

A few aspects in this comparison with the isochrones and with the
NGC~411 data, are remarkable:
\begin{enumerate}
\item The same isochrone parameters that were used to fit NGC~411,
  produce a nice fit of NGC~419 as well -- with, perhaps, just minor
  adjustments being required in the adopted extinction. Indeed, we
  have observed that a slightly better reproduction of the RC and MSTO
  loci would have been obtained with a slightly larger value of \av\
  (namely 0.28 instead of 0.25~mag) for NGC~419; however, for the sake
  of clarity in the comparison, we still adopt the same extinction
  value as for NGC~411, in all our plots.
\item The extent of the eMSTO is very similar in NGC~411 and NGC~419.
  Possibly there are differences at the bottom part of the eMSTOs,
  corresponding to the oldest stellar ages, which is also the CMD
  region in which binaries may have the largest effect in the CMD
  features. It is hard to tell whether these differences come from the
  different level of crowding between the two clusters, or if they are
  due to the two clusters spanning a different age interval.
\item Although the main concentration of RC stars is in very much the
  same CMD position for the two clusters, the SRC is clearly more
  extended towards fainter magnitudes in NGC~419 than in NGC~411. In
  particular, the NGC~419 CMD presents a clear concentration of SRC
  stars at $\fvw=20.3$ and $\fvw\!-\!\fiw=1.35$, which does not appear
  in NGC~411.
\end{enumerate}

It suffices to look at the youngest isochrone in the comparison -- in
this case the 1.45~Gyr one, in blue -- to explain this more extended
SRC in NGC~419. As thoroughly discussed in \citet{Girardi_etal09},
this 0.5~mag faint extension of the RC is the undeniable signature of
stellar cores that were just massive enough to ignite helium before
the onset of electron-degeneracy. Although these conditions can be
produced by means of interacting binaries of slightly older ages
\citep{Yang_etal11}, an easy way of producing these stars {\em in
  copious numbers} is simply that of assuming that slightly more
massive (and younger) stars are present in the cluster.  In the case
of the comparison between NGC~411 and NGC~419, the difference in the
red clump morphology can be explained by NGC~419 having a youngest
population which is just 60~Myr younger than the youngest population
is NGC~411.

Figure~\ref{fig_lfrc} compares the luminosity function (LF) of RC
stars between the two clusters, after scaling them to the same total
number as in NGC~411.  The similarity between the two LFs is
remarkable, but with one significant difference: the deficit of SRC
stars in NGC~411, at $\fvw=20.3$. A two-sample Kolmogorov-Smirnov (KS)
test reveals a probability of 3~\% that these distributions are drawn
from the same intrinsic one. However, if we correct the NGC~419 data
points by an extinction of just $\Delta A_V\simeq0.04$~mag, the same
KS turns out to provide significantly higher probabilities, as high as
97~\%. Therefore, the difference between the RC morphologies cannot be
regarded as statistically significant.

Figure~\ref{fig_lfms} instead compares the eMSTO region between the
two clusters. Following the same kind of analysis as performed in
\citet{Goudfrooij_etal11a} for LMC clusters, we draw an approximate
sequence of ages across the eMSTOs, by means of the inclined
parallelogram shown in the plot.  This is almost identical to the one
used by \citet{Goudfrooij_etal11a}, but shifted to significantly bluer
colours so as to take into account the overall shift of the CMD, due
to the lower SMC metallicities\footnote{In LMC clusters with eMSTOs,
  like NGC~1751, NGC~1783, NGC~1806, and NGC~1846, the eMSTOs appears
  at colours $\fvw\!-\!\fiw\simeq0.6$
  \citep[see][]{Goudfrooij_etal11a}.}.  This parallelogram avoids as
much as possible the crossing of isochrones of different ages, and the
presence of apparent binaries, that occurs at the top-right extremity
of the eMSTO region. Stars are counted across this area of the CMD, in
the small boxes as indicated, giving origin to the ``pseudo-age''
sequences plotted in the bottom panel.

Here, the similarity between the eMSTOs in the two clusters is again
evident: both comprise very much the same total interval of colours
and hence pseudo-ages, and the same peak at
$\fvw\!-\!\fiw\!\sim\!0.3$. The oldest bin for which the star counts
are significant, corresponds to ages of about 2.2~Gyr, whereas the
youngest one corresponds to ages of about 1.5~Gyr. There is also a
subtle difference between the two histograms: NGC~419 contains
relatively more older MSTO stars than NGC~411. At first sight, the
effect seems to be significant: a two-sample KS test indicates a
probability of $1.2$~\% that the two distributions are drawn from the
same intrinsic one. However, when correcting the NGC~419 data by a
small relative extinction w.r.t.\ NGC~411 of the order of $\Delta
A_V\simeq0.04$~mag, we find that the null hypothesis -- that both
distributions of pseudo-ages follow from the same intrinsic one --
reaches probability values as high as $\sim\!46$~\%.

Although these comparisons point to a great similarity between the
CMDs of NGC~411 and NGC~419, it is clear that any quantitative
conclusion about their {\em differences} has to wait for a more
detailed analysis, that takes into account all the possible (and
likely) small differences between the cluster distances, extinctions,
and metallicities, as well as the different photometry.

\section{Summary and concluding remarks}
\label{conclu}

Based on the new HST observations with the WFC3/UVIS camera, we report
the discovery of eMSTOs in the SMC intermediate-age cluster NGC~411.
The same cluster presents a slightly-broadened RC, compatible with the
presence of stars that, after their main sequence, developed a core
mass just massive enough to ignite helium in non-degenerate
conditions.

This is, after NGC~419 \citep{Glatt_etal09, Girardi_etal09}, just the
second SMC cluster to present such features in their CMDs. eMSTOs have
been uncovered in 13 clusters of similar ages in the LMC \citep[][and
references therein]{Milone_etal08, Piatti13, Keller_etal12}, among
which at least 6 do also seem to present a dual RC
\citep{Girardi_etal09}. The smaller number of such clusters in the SMC
is expected, because the SMC contains significantly fewer
intermediate-age and populous clusters than the LMC. Moreover, just a
handful of them has been observed with high resolution by HST. Present
observations seem to indicate that eMSTOs are as frequent an
occurrence in the SMC clusters as in the LMC. Whatever is causing such
phenomenon, it happens as well for all metallicities in the range
$-1.0\la\feh\la-0.4$.
 
\citet{Keller_etal11,Keller_etal12} find that all LMC clusters with
eMSTOs are located in a well-defined region of the core radius--age
diagram, with $R_{\rm c}\ga3.5$~pc and ages between 1 and 2~Gyr (see
figure~4 in \citealt{Keller_etal12}). With mean ages of $\sim1.66$~Gyr
and core radii of 4.2~pc and 4.5~pc, respectively, both NGC~411 and
NGC~419 fall well within the same region of this diagram, although
close to the upper age limit at which eMSTOs become difficult to
resolve.  It will be quite interesting to check if the same trends
persist when data for more SMC clusters become available.

Another result is that, in terms of their global CMD features, NGC~411
and NGC~419 are nearly perfect twins, despite the former being about 4
times less massive and populous than the latter. After having
transformed NGC~419 observations -- partially obtained with ACS/WFC --
to the same filter system as NGC~411, we have observed just minor
differences between their CMDs, which could be partially attributed to
minor offsets in their extinction values, small inadequacies in the
transformation equations, or to the different photometric conditions
in both clusters -- the NGC~419 core being significantly more crowded.
The differences are essentially: (1) NGC~411 appears to have a deficit
of SRC stars as compared to NGC~419. This deficit could be explained
either by the small number statistics, or by simply assuming that the
youngest episode of star formation in NGC~419 is about 60~Myr younger
than in NGC~411. (2) The eMSTO in NGC~419 is slightly more weighted
towards older ages than in NGC~411. Both differences, however, become
much less significant when the NGC~419 data is corrected by a
difference in $V$-band extinction of $\Delta A_V\!\simeq\!0.04$~mag
w.r.t.\ NGC~411.

The fact that both NGC~411 and NGC~419 clusters have nearly the same
mean age was already noticed by \citet{Rich_etal00}, using previous
WFPC2 observations. They also identify a third cluster with apparently
the same age (namely NGC~152), and suggest a major episode of cluster
formation in the SMC. Our present observations confirm this picture,
adding the evidence that NGC~411 and NGC~419 have some more properties
in common than simply their mean age.

In the context of the present ongoing discussion about the origin of
eMSTOs, the present observations are intriguing. If we assume that the
only factor causing eMSTOs and dual RCs in these clusters are their
extended histories of star formation, we are compelled to conclude
that NGC~411 continued forming stars for a period almost as long as
NGC~419, despite the latter being a much more massive cluster (with
about 4 times the NGC~411 present total mass). The less pronounced SRC
in NGC~411 is the only indication that it may have stopped its
internal star formation at earlier ages than NGC~419.  A modest 60~Myr
earlier halt in SFH (as compared to the total $\sim\!700$~Myr) would
suffice to explain the difference.

Such long periods of star formation, occurring in two isolated
clusters in a very similar way despite their very different present
masses, is puzzling. One possibility is that the NGC~411 total mass
was significantly higher in the past, so that {\em both} clusters had
high enough escape velocity while the star formation was occurring in
their centres. This aspect will be investigated in forthcoming papers.

On the other hand, our observations do not fit easily on the
alternative explanations for the eMSTOs and dual RCs, that have been
advanced by other authors. Indeed, two other possibilities are to be
considered here:
\begin{enumerate}
\item A spread in rotation among coeval stars, as suggested by
  \citet{BastiandeMink09}: As demonstrated by \citet{Girardi_etal11},
  this effect is unlikely to cause eMSTOs, just because rotation
  extends the main sequence lifetimes. When isochrones are built from
  ``rotating'' evolutionary tracks, the increase in lifetimes largely
  compensates for the redward extension of the tracks due to rotation.
  The net effect is that coeval populations with and without rotation
  present a very modest spread in their turn-offs. Note that
  \citet{Li_etal12} have recently repeated the same mistake of
  ignoring the effect of rotation in lifetimes. At present, such a
  ``picture'' for the formation of eMSTOs is simply missing any
  convincing demonstration that it may cause eMSTOs in {\em coeval}
  clusters\footnote{ Such a demonstration has to be done using either
    coeval isochrones or simulated CMDs for rotating and non-rotating
    stars, as done in \citet{Girardi_etal11}, and not ``corrections
    based on evolutionary tracks'', as done in \citet{BastiandeMink09}
    and \citet{Li_etal12}.}. On the other hand, recent observations by
  \citet{Platais_etal12} found that rotation causes just a very modest
  broadening of the MSTO in the intermediate-age open cluster Tr~20,
  in agreement with the model predictions by \citet{Girardi_etal11}.
\item The presence of interacting binaries, that could contribute to
  eMSTOs and dual red clumps as shown by \citet{Yang_etal11}: One main
  difficulty in this case is in explaining the high frequency (about
  15\% of all RC stars; \citealt{Girardi_etal09}) of SRC stars in a
  cluster such as NGC~419, which would require an extremely high
  fraction of interacting binaries. The fraction of non-interacting
  binaries in NGC~419, as detected from the sequence parallel to the
  cluster main sequence, is a modest $\simeq\!18$\% (limited to the
  interval of mass ratios between $\sim0.7$ and 1,
  \citealt{Rubele_etal10}).  So, in order to explain NGC~419
  observations with binaries only, we would have to assume it contains
  nearly as many interacting binaries than non-interacting ones, which
  as far as we know, would be an unprecedented and very challenging
  observation.
\end{enumerate}

Overall, our observation of a very extended MSTO in NGC~411 -- almost
as extended as in NGC~419 -- adds a puzzle to an already confusing
scenario. Although we are favouring the Occam's Razor picture of
extended SFHs occurring inside these clusters -- simply because it has
been shown to work in a quantitative way \citep[see
][]{Girardi_etal09,Rubele_etal10, Rubele_etal11, Rubele_etal13} -- we
have to admit that the similarity of such CMDs in clusters of very
different mass is intriguing. We are confident that the solutions for
these issues will come from more observations of intermediate-age LMC
and SMC clusters, together with even more accurate modelling of their
CMDs.  We are going to pursue both ways in upcoming papers.

\section*{Acknowledgements}
We are grateful to Jay Anderson for sharing his ePSF program.
The data presented in this paper were partially obtained from the
Multimission Archive at the Space Telescope Science Institute (MAST).
Support for {\it HST\/} Program GO-12257 was provided by NASA through
a grant from the Space Telescope Science Institute, which is operated
by the Association of Universities for Research in Astronomy, Inc.,
under NASA contract NAS5--26555. We thank the partial support from
contract ASI-INAF I/016/07/0.
%
%

%
\label{lastpage}
\end{document}